\documentclass[12pt]{iopart}

\newcommand{\ud}{\mathrm{d}}
\newcommand{\RacS}{\sqrt{s}}
\newcommand{\dedx}{\ud E/ \ud x}
\newcommand{\ks}{\mathrm{K}^{0}_{\mathrm{s}}}
\newcommand{\kp}{\mathrm{K}^{+}}
\newcommand{\km}{\mathrm{K}^{-}}
\newcommand{\lam}{\Lambda}

\usepackage{graphicx}

\begin{document}

\title[Study of strange particle production in pp collisions with the ALICE detector]{Study of strange particle production in pp collisions
 with the ALICE detector}

\author{H\'el\`{e}ne Ricaud$^{1}$, Alexander Kalweit$^{1}$ and Antonin Maire$^{2}$ (for the ALICE Collaboration)}

\address{$^{1}$Technical University of Darmstadt,
Institut fuer Kernphysik,
Schlossgartenstrasse 9, D-64289 Darmstadt, Germany}
\address{$^{2}$Institut Pluridisciplinaire Hubert Curien, 23 rue du Loess, F-67037 Strasbourg, France}

\ead{H.Ricaud@gsi.de}
\begin{abstract}
ALICE is well suited for strange particles production studies since it has very good reconstruction capabilities in the low transverse momentum ($p_{t}$) region and it also allows to extend the identification up to quite high $p_{t}$. Charged strange mesons ($\kp$, $\km$,) are reconstructed via energy loss measurements whereas neutral strange mesons ($\ks$) and strange hyperons ($\lam$, $\Xi$, $\Omega$) are identified via vertex reconstruction. All these particles carry important information: first, the measurement of production yields and the particle ratio within the statistical models can help to understand the medium created and secondly the dynamics at intermediate $p_{t}$ investigated via the baryon over meson ratio ($\lam / \ks$) allows a better understanding of the hadronization mechanisms and of the underlying event processes. We present these two aspects of the strange particles analysis in pp collisions using simulated data.
\end{abstract}

\maketitle

\section{Introduction}
Over the past few years, strange particle production has been largely studied and it has been proven to be a good tool to investigate the Quark Gluon Plasma (QGP) which can be created in high energy density conditions such as relativistic heavy-ion collisions.  Strange particle analysis will be important at LHC to characterize the medium created in the future Pb-Pb collisions but it can also give very useful information to understand pp collisions. These elementary collisions are expected to be a baseline for heavy-ion physics but they can also be interesting in themselves as we will show.

\smallskip

In the second section, we describe the ALICE detector and its particle identification abilities (see also~\cite{BelikovProceeding}). The third section will be dedicated to two aspects of the physics we can extract from strange particle analysis in pp collisions. The first one is the chemical composition analysis done within statistical models. These models have been very successful in describing hadron yields in central heavy-ion collisions. We present here their applicability in small systems like elementary collisions and their predictions at the top LHC energy for pp collisions. We focus then on the dynamics at intermediate $p_{t}$ via the measurement of the ratio $\lam / \ks$ versus $p_{t}$.

\section{The ALICE detector and the particle reconstruction}

The design of the ALICE detector is ideal for particle reconstruction in the soft physics sector. The reconstruction of strange particles is performed inside the so called central barrel of ALICE which is embedded in a large solenoidal magnet and covers polar angles from 45\ensuremath{^\circ} to 135\ensuremath{^\circ} over the full azimuth. Tracking in the central barrel is mainly divided into the Inner Tracking System (ITS) and the Time Projection Chamber (TPC), the main tracking detector of ALICE. 

\smallskip

The particle identification (PID) capabilities of ALICE are remarkable and cover a large part of the phase space by using all known experimental techniques. Several detectors can be used for PID independently or in combination. In addition to tracking
the TPC can serve as a PID detector in the non-relativistic and relativistic rise regions of the Bethe-Bloch parametrization. via the energy loss measurements ($\dedx$). The four outer layers of ITS have analog PID capabilities via $\dedx$ measurement in the non-relativistic region. This technique can be used for $\ks$, $\lam$, $\Xi$ and $\Omega$ analysis but also secondary vertex topology identification such as from $\ks$ and $\lam$ can help extend PID capabilities to higher $p_{t}$. 

\smallskip

\begin{figure}[!h]
   \begin{minipage}[t]{.48\linewidth}
    \centering\includegraphics[width=7cm, height=6.5cm]{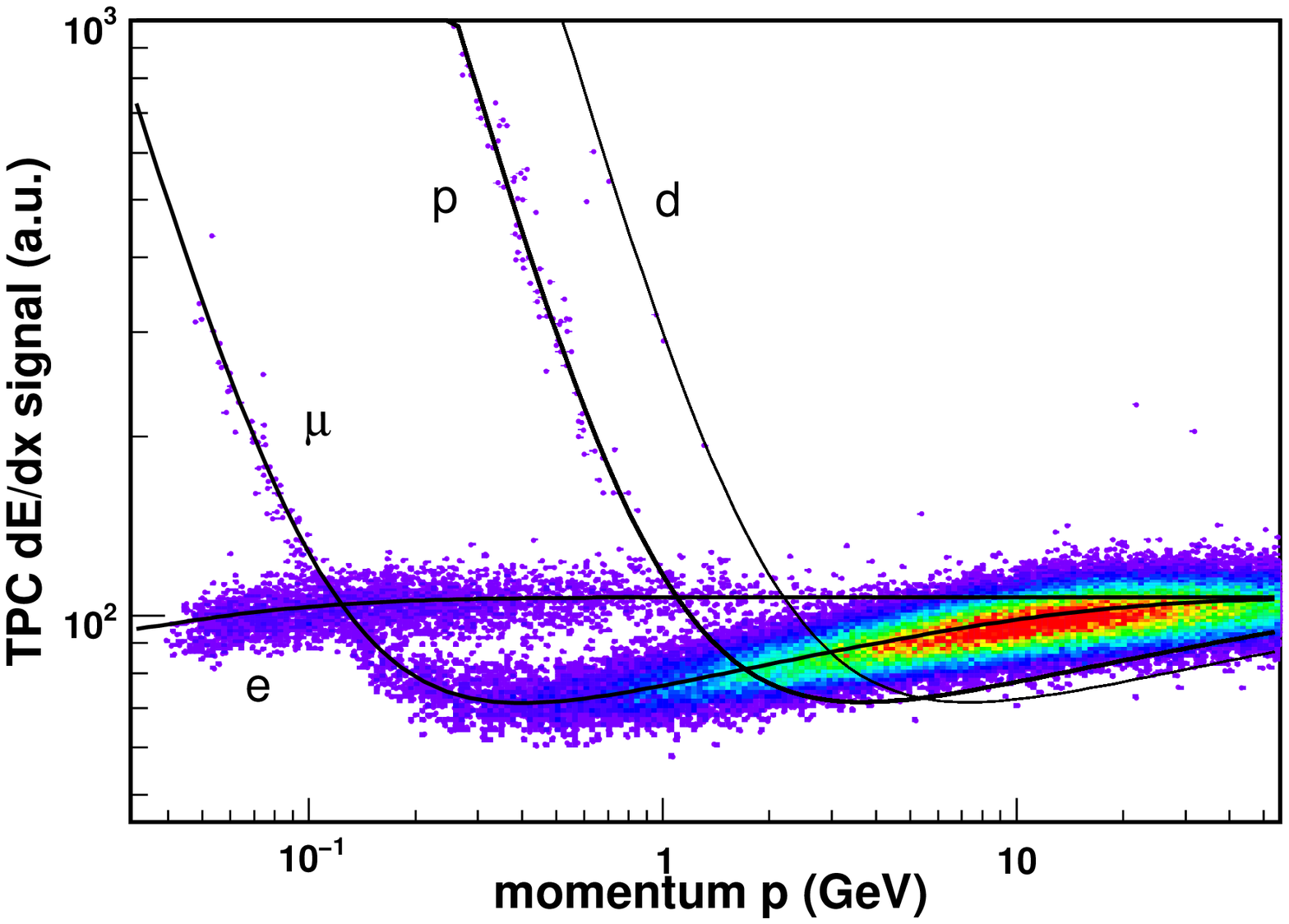}
    \caption[]{\label{FigdedxCosmis} Energy loss in the TPC versus momentum measured with cosmics.}
   \end{minipage} \hfill
   \begin{minipage}[t]{.48\linewidth}
    \centering\includegraphics[width=7.5cm, height=6.5cm]{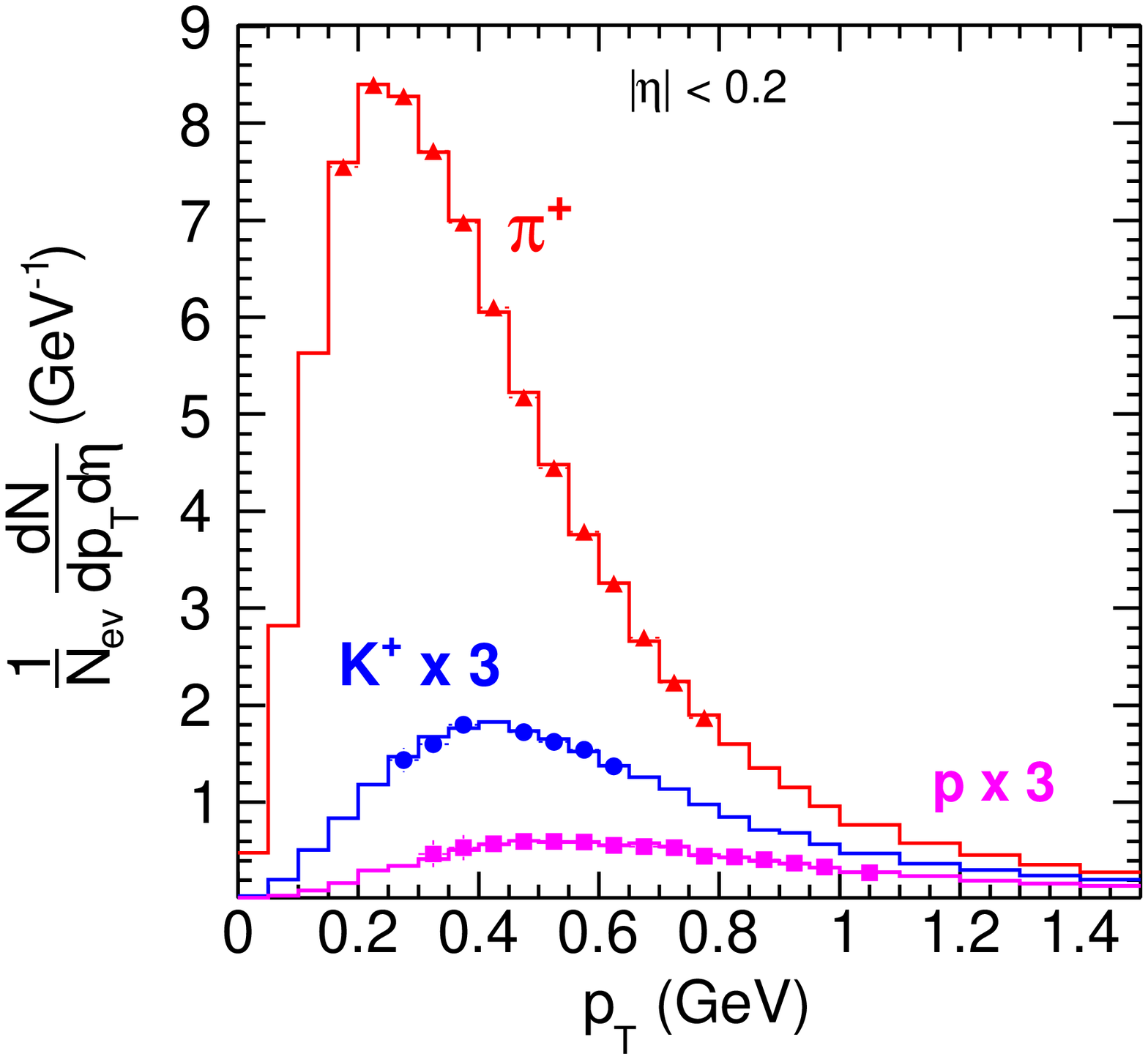}
    \caption[]{\label{FigSpectraKaonsSimData} $\kp$, $\pi^{+}$ and proton $p_{t}$ spectra at mid-rapidity in simulated data.}
   \end{minipage}
\end{figure}

The ALICE TPC provides excellent particle identification in the low $p_{t}$ region which allows us to extract charged kaons spectra for $p_{t}<1.5~\mathrm{GeV/\textit{c}}$ by deconvolution of several Gaussians in each $p_{t}$ interval. It will be possible to obtain $\kp$ and $\km$ spectra at higher $p_{t}$ with the help of the Time of Flight detector and the PID in the relativistic rise with the TPC. We present here in Fig.~\ref{FigdedxCosmis} the $\dedx$ histogram obtained from high statistics cosmics data. The $\kp$ and $\km$ spectra from simulated data are presented in Fig.~\ref{FigSpectraKaonsSimData}. The particle ratios such as $\kp / \pi^{+}$, $\kp / \km$, $\mathrm{p} / \pi^{+}$ will be studied as a function of $p_{t}$ and multiplicity in pp collisions.

\smallskip

The multi-strange particles, $\Xi$ and $\Omega$, are reconstucted with the help of the ITS and TPC detectors via their decay products. We estimate that, at $\RacS=10~\mathrm{TeV}$, we will need 100 M minimum bias events to obtain $\Xi$ spectra up to $p_{t}=7~\mathrm{GeV/\textit{c}}$ with good uncertainties. We could however have access to spectra up to $p_{t}=4~\mathrm{GeV/\textit{c}}$ with 10\%  of statistical uncertainties for each point with 2 M minimum bias events.

\smallskip

$\ks$ and $\lam$ are also reconstructed topologically in the TPC and ITS. In order to reduce the background and purify the sample, it is possible to use the TPC PID to better identify the decay products but we have shown that we can extract the signal even without this help. The signal extraction is also straightforward without any further selection optimization. It is however impossible to reconstruct $\ks$ and $\lam$ below $p_{t}=0.2~\mathrm{GeV/\textit{c}}$ and $p_{t}=0.5~\mathrm{GeV/\textit{c}}$ respectively. Our analysis showed that the comfortable situation to extract $p_{t}$ spectra will be to have 1 M events but it will be feasible to reach $p_{t}=3.0~\mathrm{GeV/\textit{c}}$ in the $\lam$ spectrum only with 200 K events.

\section{Physics at LHC with strange particles}

\subsection{Perspectives of statistical model analysis at LHC}
Particle production calculated within the statistical models is quantified by several thermal parameters: temperature, volume and a set of chemical potentials related to charge conservation. In pp collisions, strangeness conservation is treated canonically. However, SPS data have shown a stronger suppression of strange particles yields than expected in the canonical model~\cite{Becattini:2005xt}. As a consequence, an additional suppression effect has to be added. This can be done by using the $\gamma_{S}$ fugacity factor to take into account deviations of strange particles abundance from their chemical equilibrium distribution value~\cite{StatModelGammaS} or by using the $R_{c}$ parameter to control the radius of the sub-volume within which strangeness is conserved exactly. In the following, we will focus on the second option. A statistical model that quantifies the extra strangeness suppression by this strangeness correlation volume has been formulated in~\cite{KrausProceeding,StatModelRc1,StatModelRc2}.

\smallskip

The extrapolation of the parameters of the statistical models from SPS to LHC is quite straightforward for temperature and chemical potentials. No temperature variation is expected between RHIC and LHC whereas the baryonic potential $\mu_{B}$ should decrease. However, the value of the $R_{c}$ parameter remains largely uncertain at LHC~\cite{PartProdppLHC}, either it saturates and therefore does not vary with the collision energy or it keeps increasing. Strange particles and above all multi-strange particle ratios are shown to be very sensitive to the value of $R_{c}$. For large values of this parameter, we clearly observe that particles ratios are very close to the Grand-Canonical description used for large systems like heavy-ion collisions~\cite{PartProdppLHC}.

\subsection{Baryon over meson ratio}

In RHIC central heavy-ion collisions, quark coalescence has been suggested as a possible hadronization mechanism which explains qualitatively the amplitude of the mid-rapidity baryon/meson ratio at intermediate $p_{t}$. In pp collisions, quark coalescence is clearly not favoured due to the low phase space density. The interest of strange particles such as $\lam$ and $\ks$ in this analysis is due to their topological reconstruction techniques that allow an identification over a large $p_{t}$ range. 

\subsubsection{Baryon over meson ratio in elementary collisions: $\lam / \ks$ at UA1}
\begin{figure}[!h]
\begin{center}
  \includegraphics[width=9cm,height=7.7cm]{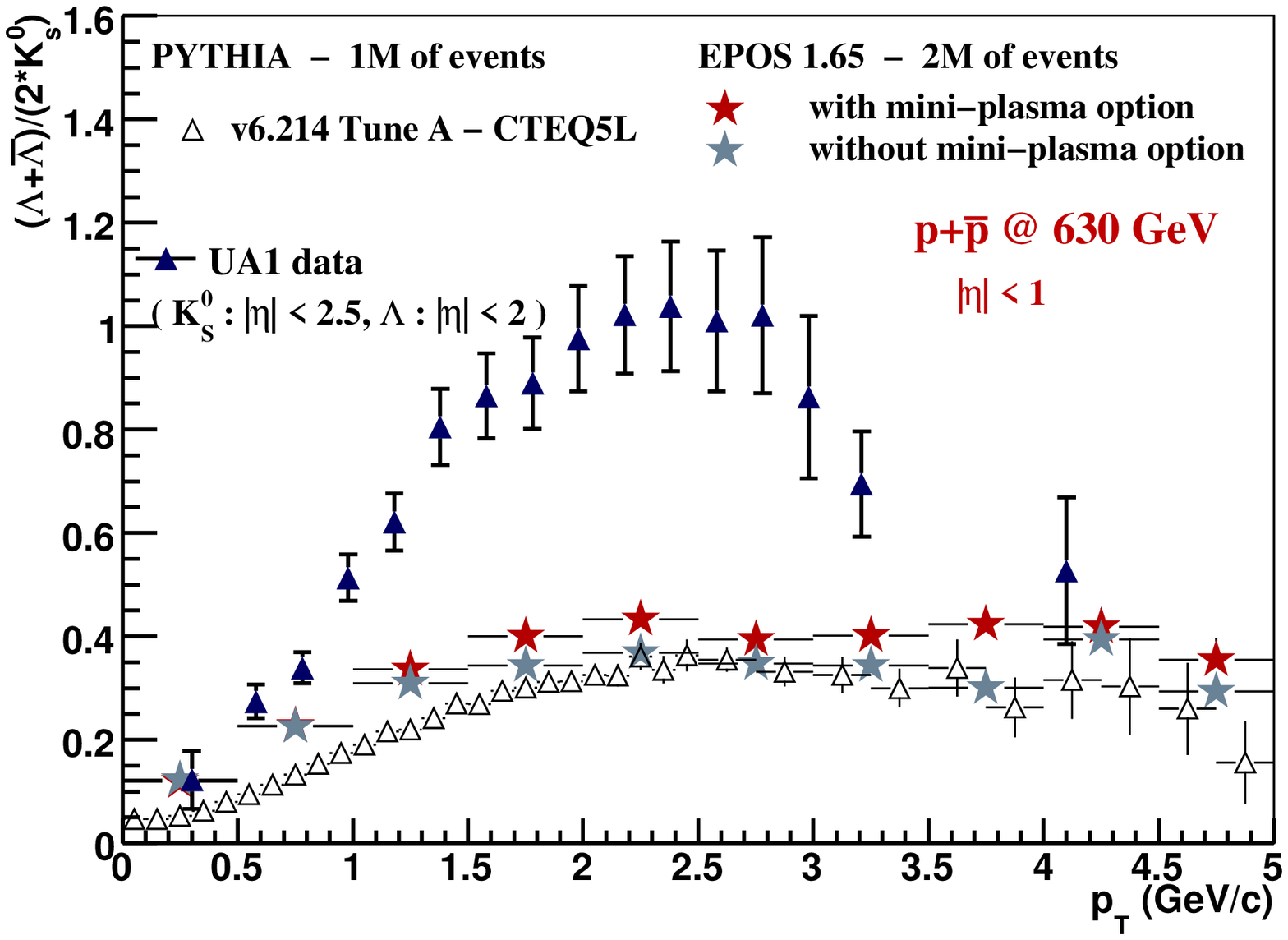}
  \caption[]{\label{FigUA1Ratio}$\lam / \ks$ in $\mathrm{p}+\bar{\mathrm{p}}$ collisions at $\RacS=630~\mathrm{GeV}$ at mid-rapidity (UA1 experiment) compared to PYTHIA and EPOS simulations.}
\end{center}
\end{figure}

In pp collisions at RHIC, the $\lam / \ks$ ratio showed a maximum amplitude of ~0.6, well below the one obtained in central heavy-ion collisions. This value is consistent with the one obtained at HERA in the $\RacS=319~\mathrm{GeV}$ e+p collisions where the hadronization is modeled by fragmentation~\cite{RefHERA}. However as it is shown in Fig.~\ref{FigUA1Ratio}, the $\lam / \ks$ ratio in UA1 $\mathrm{p}+\bar{\mathrm{p}}$ collisions at $\RacS=630~\mathrm{GeV}$ is close to unity. The baryon over meson ratio reaches even higher values at higher energy. We cannot a priori invoke coalescence in pp collisions to explain this amplitude but on the other hand none of the two models, PYTHIA and EPOS, can reproduce the experimental data as shown in Fig.~\ref{FigUA1Ratio}.

\subsubsection{Baryon over meson ratio: predictions of PYTHIA and EPOS at LHC}
\label{SectionPredictionLHC}
\begin{figure}[!h]
   \begin{minipage}[t]{.5\linewidth}
    \centering\includegraphics[width=7.5cm, height=6.5cm]{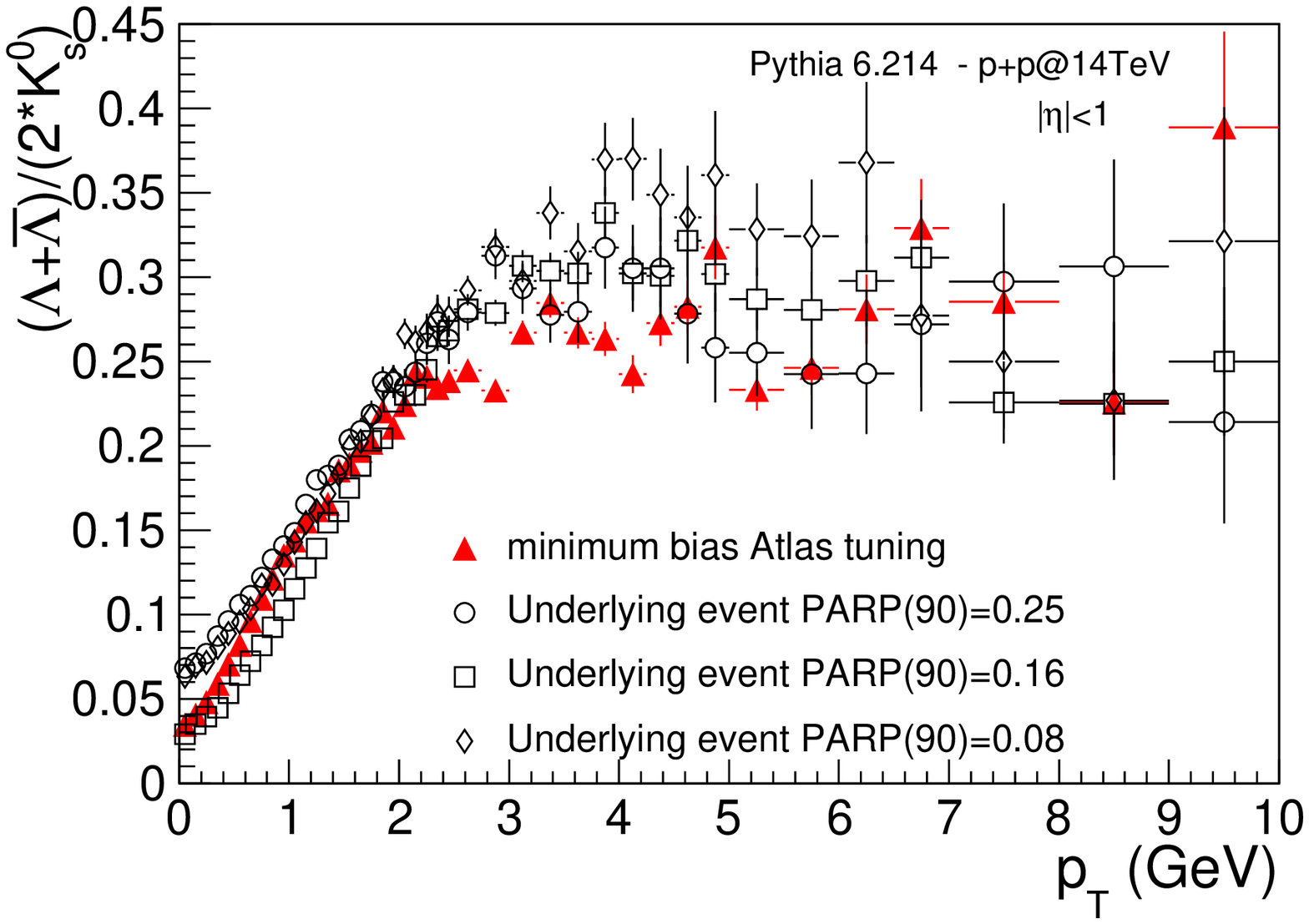}
    \caption[]{\label{FigPythiaLHC} Predictions of PYTHIA 6.214 for pp collisions at $\RacS=14~\mathrm{TeV}$.}
   \end{minipage} \hfill
   \begin{minipage}[t]{.5\linewidth}
    \centering\includegraphics[width=7.5cm, height=6.5cm]{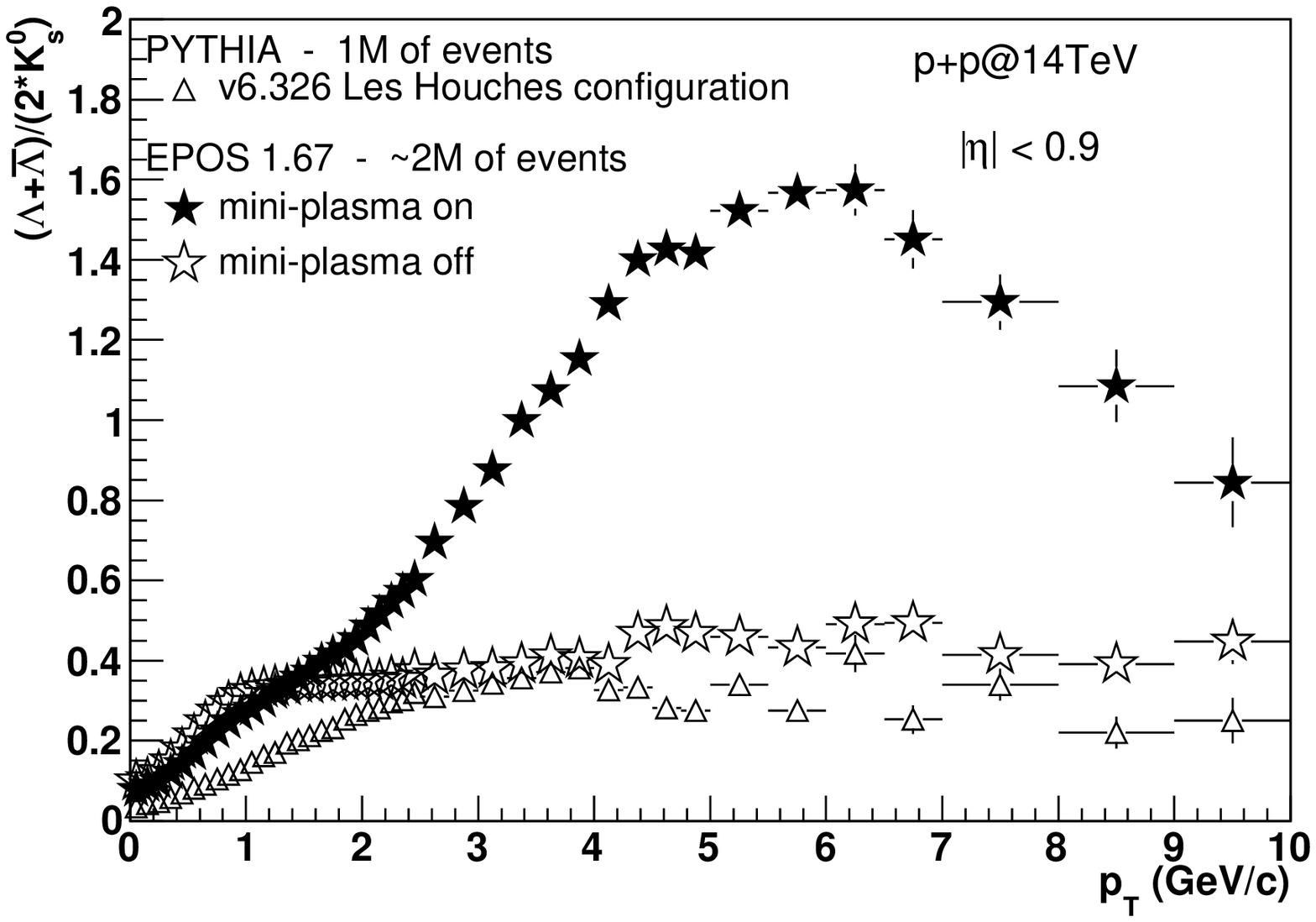}
    \caption[]{\label{FigEPOSLHC} Predictions of EPOS 1.65 for pp collisions at $\RacS=14~\mathrm{TeV}$ compared with PYTHIA.}
   \end{minipage}
\end{figure}

In order to understand the mechanisms that take place in pp collisions and that lead to such a high amplitude of the baryon over meson ratio, we have studied two different models, PYTHIA and EPOS, and their predictions at LHC energies. Using PYTHIA~\cite{RefPYTHIA}, a model based on the Lund fragmentation, requires to extrapolate all the parameters from TEVATRON to LHC. To do so the so-called ATLAS tune has been used by ALICE~\cite{RefAtlasTuning}. This set of parameters leads to a very low $\lam / \ks$ ratio at $\RacS=14~\mathrm{TeV}$, considerably lower than the UA1 data, but in line with the PYTHIA predictions for the UA1 energies (see Fig.~\ref{FigPythiaLHC} and~\cite{HotQuakHippolyte}). With the increase of the collision energy, the probability of multiple-parton interactions  is likely to increase. We have therefore modified the ATLAS set of parameters in order to better describe the so-called underlying event, which includes
all phenomena other than the hard scattering process, such as the multiple parton interactions. The values of the parameter PARP(90) that controls the multiple interaction rate (see Fig.~\ref{FigPythiaLHC}) have been chosen according to PYTHIA authors' suggestions. Increasing the multiple interaction rate helps to increase the baryon over meson ratio so it seems to go in the right direction but a factor 2 at least is still missing (see Fig.~\ref{FigPythiaLHC}). We have then investigated EPOS, a multiple interaction model based on Pomeron exchanges~\cite{RefEPOS}. The specificity of EPOS is that it considers the possibility that collective phenomena take place in pp collisions (mini-plasma). As we have seen on Fig.~\ref{FigUA1Ratio}, this mini-plasma option has no effect at $\RacS=630~\mathrm{GeV}$. However, the effect is dramatic at $\RacS=14~\mathrm{TeV}$, it leads to a very strong increase of the baryon over meson ratio (see Fig.~\ref{FigEPOSLHC}). The high amplitude of the ratio obtained with the mini-plasma option is due to two combined effects: a stronger increase of the $\lam$ yield than of the $\ks$ one and a change in the shape of the $p_{t}$ spectra: it is strongly shifted to higher $p_{t}$ in the case of the baryon. If LHC data show such a high amplitude of the $\lam / \ks$ ratio, it could be interpreted, according to EPOS, as the manifestation of collective phenomena in pp collisions.

\section{Conclusion}
We have discussed several aspects of the anticipated strange particle analysis in pp collisions at LHC energies. Strange particles are an interesting tool to investigate the nature of the system created in pp collisions. They could reveal, at the LHC energies, that the canonical suppression might diminish, which means strange particle yields could reach the value obtained in heavy-ion collisions. Strange particle ratios could also prove the presence of collective phenomena. Both aspects that have been presented, production yield and dynamics at intermediate $p_{t}$, come to the same conclusion:  they could lead us to revise our view of pp collisions at very high energy, which would raise a lot of questions concerning the use of pp collisions as a baseline but would certainly open a perspective on new physics in pp collisions.

\end{document}